\documentclass[twocolumn,showpacs,preprintnumbers,amsmath,amssymb,floatfix,prb]{revtex4}

\usepackage{graphicx}
\usepackage{dcolumn}
\usepackage{bm}
\usepackage{indentfirst}

\usepackage{epsfig}
\usepackage{graphics}

\usepackage{amssymb}
\usepackage{amsmath}
\usepackage{amsthm}

\usepackage[latin1]{inputenc} 
\usepackage{setspace}
\usepackage{makeidx}
\usepackage[mathscr]{eucal}
\usepackage{mathrsfs}
\usepackage{latexsym}
\usepackage{eufrak}
\usepackage{color}

\begin{document}

\title{
Band offsets and density of Ti$^{3+}$ states probed by X-ray
photoemission on LaAlO$_{3}$/SrTiO$_{3}$ heterointerfaces and their
LaAlO$_{3}$ and SrTiO$_{3}$ bulk precursors

}

\author{G. Drera $^{1}$, G. Salvinelli$^{1}$,  A. Brinkman$^{2}$, M. Huijben$^{2}$, G. Koster$^{2}$, H. Hilgenkamp$^{2}$,
G. Rijnders$^{2}$, D. Visentin$^{1}$, and L. Sangaletti$^{1}$ \\}

\affiliation{%
$^{1}$Interdisciplinary Laboratories for Advanced Materials Physics
and Dipartimento di Matematica e Fisica, Universit\`{a} Cattolica,
via dei Musei 41, 25121 Brescia (Italy)}%

\affiliation{$^{2}$ MESA+ Institute for Nanotechnology, University of Twente, PO Box 217, 7500 AE Enschede, The Netherlands}%

\date{\today}

\begin{abstract}
A set of $LaAlO_3/SrTiO_3$ (LAO-STO) interfaces has been probed by
x-ray photoemission spectroscopy in order to contrast and compare
the effects of LAO overlayer thickness and of the growth conditions
on the electronic properties of these heterostructures. These
effects are tracked by considering the band-offset and the density
of Ti$^{+3}$ states, respectively. It is shown that the dominant
effects on the local electronic properties are determined by the
O$_{2}$ partial pressure during the growth. In particular, a low
P(O$_{2}$) yields Ti$^{+3}$ states with higher density and lower
binding energy as compared to the sample grown at high P(O$_{2}$) or
to the bare STO reference sample. Band offset effects are all below
about 0.7 eV, but a careful analysis of Ti 2p and Sr 3d peaks shows
that valence band offsets can be at the origin of the observed peak
width. In particular, the largest offset is shown by the conducting
sample, that displays the largest Ti 2p and Sr 3d peak widths.

\end{abstract}

\pacs{Valid PACS appear here}
\maketitle

\section{Introduction}
Lanthanum aluminate (LaAlO$_3$, LAO in short) and strontium titanate
(SrTiO$_{3}$, STO) are formally band insulators, as they are
closed-shell compounds (4f$^0$ for LAO and 3d$^0$ for STO), with a
band gap of 5.4 and 3.2 eV, respectively. These materials belong to
the perovskite group, sharing the same chemical formula (ABO$_3$)
and a similar cubic crystal structure. When a LAO-STO
heterointerface is created, a p-type heterostructure is expected if
the bulk STO is terminated with a SrO plane (hole doping), while an
n-type heterostructure should be obtained with a TiO$_2$ plane
(electron doping) termination. In the latter case, the LAO-STO
interface becomes conducting\cite{LAOSTO_ohtomo} and yields a
quasi-2D electron gas (2DEG). The transition to the metallic state
was found to be thickness dependent: the 2DEG is observed only when
the LAO capping is at least 4 u.c. (unit cell)
thick\cite{LAOSTO_science}. The main difference between LAO and STO
resides in the layer charge polarity\cite{LAOSTO_rev1}: looking at
the (001) planes, STO is a non-polar solid, since both
Sr$^{2+}$O$^{2-}$ and Ti$^{4+}$O$_{2}^{2-}$ planes are
charge-neutral, while LAO is a polar solid, as it is composed of
La$^{3+}$O$^{2-}$ and Al$^{3+}$O$_{2}^{2-}$ charged layers. The
p-type interface is thus formed by SrO - (AlO$_{2})^{1-}$ planes,
while the n-type by TiO$_{2}$ - (LaO)$^{1+}$ planes. The observed
conductivity was originally thought to be the response of the system
to the diverging potential (the so-called polar catastrophe
\cite{polarcatastrophe1,polarcatastrophe2}) created by LAO.

In non-oxide semiconductors, the relaxation of polar discontinuity
in heterointerfaces is usually achieved by an atomic reconstruction
process\cite{LAOSTO_ohtomo}, where the interface stoichiometry is
altered by interdiffusion, point defect, dislocation and in general
by a structural roughening. In oxides, the possibility of multiple
valence ions allows also an electronic reconstruction that, in
LAO-STO case, should move electrons from the surface to the empty Ti
d levels, leading to 3d$^{1}$ electronic states. A 2D lattice of
electrons in a correlated material can originate phenomena like MIT
transitions, localized magnetic moments and even superconductivity
\cite{LAOSTO_rev2, LAOSTO_rev3}; most of these effects have been
observed in LAO-STO, though not all in one sample at the same time.

In principle, both atomic and electronic reconstruction could be
present in the LAO-STO case. For example, there are many
experimental proofs of interdiffusion \cite{LAOSTO_chambers, VONK}
(with La ions drifting inside STO, a form of atomic reconstruction),
but this mechanism alone cannot be solely responsible of the
conductivity, since in principle it should be present also in the
p-type interface. On the other hand, standard polar catastrophe (an
electronic reconstruction picture) was invoked to explain the lack
of conductivity for thickness below 4 u.c, but a discrepancy exists
between the theoretically expected and the measured band bending
effects. In fact, the thickness dependence of conductivity could be
explained by a band bending effect\cite{BandBending}, induced by
polarity discontinuity: the density of states (DOS) of LAO valence
band (VB) should be shifted to higher binding energies till, for a
capping equal or major of 4 u.c., the VB maximum is superimposed to
the buried empty levels of STO. The conduction should be now
triggered by a tunneling effect from the surface to the interface.
However, a significant band bending has not yet been observed in
terms of core-level shift, while a shift of 3.2 eV (needed to span
the electronic gap in STO) should be easily observed.

Finally, the sample growth conditions deeply affect the transport
properties; an oxygen-poor growth atmosphere can induce oxygen
vacancies and thus a 3D conductivity\cite{OxyVac1,OxyVac2, OxyVac3,
OxyVac4}, while an excessively-rich one can even result in a 3D
growth and thus in a different kind of
heterostructure\cite{LAOSTO_rev2}. Depending on the O$_{2}$ partial
pressure during growth, three phases are usually identified,
\cite{LAOSTO_rev3, LAOSTO_natmat}: one dominated by oxygen vacancies
contribution (P$_{O_{2}}\simeq 10^{-6}$ mBar), one displaying
superconductivity (P$_{O_{2}}\simeq 10^{-5}$ mBar) and one
displaying a magnetic behavior (P$_{O_{2}}\simeq 10^{-3}$ mBar). It
is quite a challenging task to find a unified description of all
these phenomena.

The experimental signatures of many of the proposed models are the
LAO-STO band offset, and the density and distribution of Ti$^{3+}$
states. Recently, these states have been evidenced by soft X-ray
photoemission with photon energy tuned at the Ti 2p-3d threshold
\cite{APL_LAOSTO, KOI_2011}. Also core level XPS can provide an
indication of Ti$^{3+}$ states as reported in Ref.\cite{KOI_2011,
fuji_2011} and in hard X-ray photoelectron spectroscopy (HAXPES)
experiments \cite{LAOSTO_haxpes}. At odds with Ref.\cite{KOI_2011,
fuji_2011, APL_LAOSTO}, Ref.\cite{LAOSTO_chambers} did not report on
observed Ti$^{3+}$ features in Ti 2p core level photoemission.

As for valence band offsets (VBO), core level shifts much smaller
than those predicted by the polar catastrophe have been observed by
Takizawa \emph{et al.}\cite{fuji_2011}, Chambers \emph{et al.}
\cite{LAOSTO_chambers} and Segal \emph{et al.} \cite{segall} but it
is not yet possible to draw a consistent picture of VBO as the
results show in some cases opposite trends.

In the present study, a spectroscopic investigation of insulating
and conductive LAO-STO films is carried out by X-ray Photoelectron
Spectroscopy (XPS). We address the problem of the spectroscopic
signature of the interface effects by comparing several LAO-STO
interfaces with their LAO and STO bulk precursors. We checked to
which extent the electronic properties of LAO-STO can be described
as a weighted (i.e. thickness dependent) superposition of those of
LAO and STO, as deviations from this mere superposition should be
regarded as a signature of new electronic states arising from the
heterointerface build-up.

Following this approach, we show that the different LAO overlayer
thicknesses affect the electronic properties of the interfaces in
terms of band-offsets, though smaller that those theoretically
predicted, whereas the dominant effects on the local electronic
properties are related to the O$_{2}$ partial pressure during the
growth. This is seen by tracking the Ti$^{3+}$/Ti$^{4+}$ ratio for
Ti 2p core levels, and the Sr 3d and Ti 2p core level line widths.
Though band offsets are all below about 0.7 eV, differences are
detected among the samples and it is shown that the Ti 2p and Sr 3d
peak widths scale with band offsets, the larger values being found
for the 5 u.c. conducting sample. Finally, a careful analysis of the
Ti$^{3+}$/Ti$^{4+}$ peak area ratio based on the depth distribution
function of photoelectrons, allowed us to set a lower limit to the
density of Ti$^{3+}$ states across the interface as seen by the
present photoemission experiment.

\section{Experimental details}
The LAO-STO heterostructures (HS) have been grown by pulsed laser
deposition at the MESA$^{+}$ Institute for Nanotechnology,
University of Twente. The two n-type 3 u.c. and 5 u.c. LAO-STO
samples were grown in a P$_{O_{2}} \sim 10^{-3}$ mBar oxygen partial
pressure. In addition, an \textit{insulating} n-type 5 u.c. LAO-STO
sample, grown at 10$^{-1}$ mBar O$_{2}$ partial pressure has been
analyzed. Two reference single-crystal LAO and STO samples
terminated with the (001) surface have also been considered (MaTeck
GMBH). The sample list is reported in Table I. XPS has been used to
measure the core-level electronic structure and to evaluate the
stoichiometry of the heterostructures. The XPS data have been
collected at the Surface Science and Spectroscopy Lab of the
Università Cattolica (Brescia, Italy) with a non-monochromatized
dual-anode PsP x-ray source; the Mg k$_\alpha$ line ($h\nu$=1253.6
eV) has been used to achieve a better resolution (about 0.7 eV),
while the Al k$_\alpha$ line ($h\nu$=1486.6 eV) has been used when
the maximum probing depth was needed. The analyzer for XPS was a
SCIENTA R3000, operating in the transmission mode, which maximizes
the transmittance and works with a 30$^{°}$ acceptance angle.

In XPS, the core-level peak area of a selected layer at a depth $d$
with a thickness $t$ can be evaluated through the following formula:

\begin{equation}\label{eq_ddf}
I(E_k,\alpha)=K\cdot \int_{d}^{d+t}\Phi(E_k,\alpha,z)dz
\end{equation}

where K is a normalization constant, which includes the
photoionization cross section, the atomic density of the species and
analyzer-dependent parameters; $\Phi(E_k,\alpha,z)$ is the generical
escape probability (known as depth distribution function, DDF) of an
electron generated at a depth $z$ with a kinetic energy $E_k$ at an
angle $\alpha$ respect to the surface normal.

According to the Lambert-Beer law, the DDF function is usually
approximated with a Poisson distribution $\Phi=e^{-z/\lambda
Cos(\alpha)}$, where $\lambda$ is the inelastic mean free path
(IMFP). Even if it leads to simple analytical expressions for the
peak areas, such approximation (defined as straight line motion by
Tilinin \emph{et al.}\cite{TilinDDF}), is known to be quantitatively
wrong and, especially for the present thin overlayers, can results
in an overestimated capping thickness.

In this work, we resorted to Monte-Carlo (MC) DDF calculations, with
the algorithm described in Ref.\cite{WernerDDF}, in order to include
inelastic as well as elastic electronic scattering, in the so-called
transport approximation\cite{JablonskiTA} (TA). The photoemission
asymmetry parameters have been taken into account for each
core-level. Monte-Carlo calculations of electron trajectories have
been carried out in order to predict the XPS peak areas in LAO-STO
heterostructures, since an analytic DDF formulation\cite{TilinDDF}
can not be written for a generic multilayer sample.

\section{Results and Discussion}

\subsection{Core level and valence band
photoemission}

\begin{figure}
\begin{center}
\includegraphics[width=0.48\textwidth]{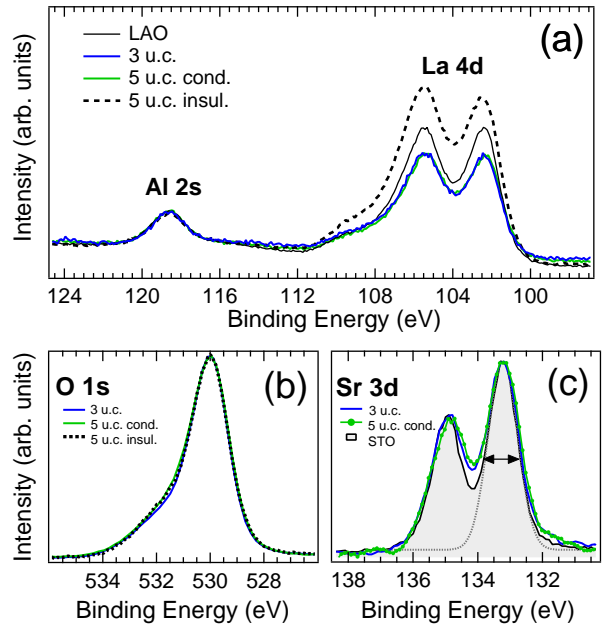}
\caption{(Color online) (a)  Al 2s and La 4d XPS spectra collected
from LAO and the LAO-STO heterostructures. Spectra have been
normalized to the Al 2s peak area. O 1s XP spectra of the three HS
(b) and Sr 3d XPS spectra of the STO single crystal and the two HS
grown at low P(O$_{2}$) (c). \label{fig_XPS_SrAlLa}}
\end{center}
\end{figure}

The La 4d  and Al 2s XPS shallow core levels are
shown in Fig.\ref{fig_XPS_SrAlLa}-a. All spectra are normalized to
the Al 2s peak. As can be observed, for the the 5 u.c. conducting
and the 3 u.c. insulating samples the La 4d core levels are
superposed, with an intensity below that of the LAO reference
crystal. In turn, the La 4d XPS core lines of the 5 u.c. insulating
sample display the largest intensity. These features are
qualitatively consistent with the results reported in
literature\cite{QIAO2011}, and suggest that the P(O$_{2}$) value has
a relevant effect on the cation stoichiometry in the LAO overlayer.
This is not unexpected, as the 5 u.c. insulating sample is known to
show a 3D growth regime rather than the layer-by-layer regime for
the other two heterostructures grown at lower P(O$_{2}$). It has
been observed \cite{QIAO2011} that among possible defects related to
a La excess with respect to Al, the formation of Al vacancies seems
to be the most likely scenario. In turn, the low La intensity of the
heterostructures grown at low P(O$_{2}$) as compared to the case of
the LAO single crystal, apparently shows a La deficiency that can be
related to La diffusion through the interface, La substoichiometry,
or both.

The O 1s spectra from the three heterostructures are shown in
Fig.\ref{fig_XPS_SrAlLa}-b. It is important to note that the three
spectra are virtually identical, in spite of the different growth
conditions and thickness of the LAO overlayer. This assures that in
all cases a similar oxygen stoichiometry can be estimated at the
surface, ruling out the possibility that changes at the interface
could be ascribed to major differences in the oxygen stoichiometry
on the surface.

\begin{figure}
\begin{center}
\includegraphics[width=0.48\textwidth]{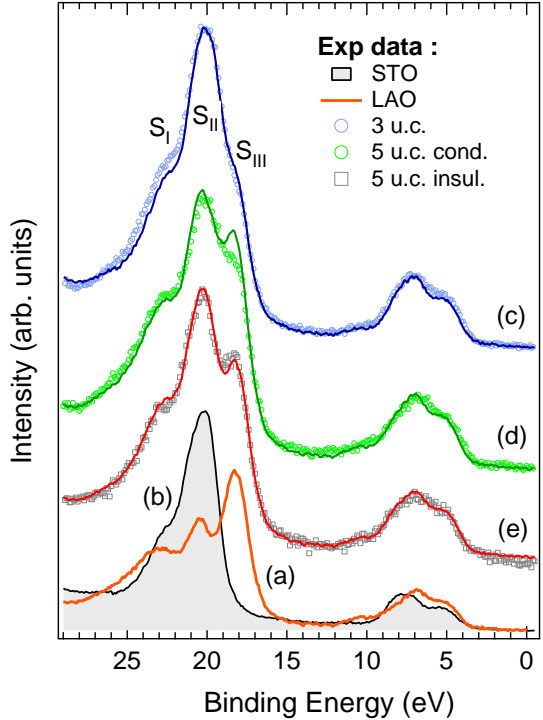}
\caption{ (Color online) XPS valence band and shallow core level
spectra of LAO and STO single crystals (a and b) and LAO-STO
heterostructures (c-e). The LAO-STO valence bands (c-e) have been
fitted with a linear combination of the LAO and STO spectra shown in
(a-b). \label{fig_VB}}
\end{center}
\end{figure}

In Fig.\ref{fig_VB} the valence band (VB) spectra, collected with
the Mg k\textnormal{$_{\alpha}$} x-ray source, are shown. The
shallow core levels are labeled as S$_{I}$, S$_{II}$ and S$_{III}$.
The O 2s states mostly contribute to the S$_{I}$ peak, Sr 4p is an
unresolved doublet below S$_{II}$ peak and La 5p is split in
S$_{II}$ and S$_{III}$ peaks (the spin orbit energy separation is
about 2.4 eV, as detected in, e.g., in La$_2$O$_3$ \cite{La2O3}). It
is possible to describe the LAO-STO spectra as the linear
combination of the single-crystal spectra. Starting from the LAO and
STO spectra as displayed in Fig.2, the VB of the three LAO-STO
samples have been calculated as a linear combination of the bulk
precursor spectra where two fitting parameters have been considered,
i.e. the energy shift of the LAO VB spectrum with respect to the STO
VB spectrum, and the relative integrated intensity of these two VB
spectra. The results of this procedure are shown in
fig.\ref{fig_VB}, thin lines. In all cases, there is a slight
difference (below 0.2 eV) among the energy shifts resulting from the
best fit of the three LAO-STO interfaces (Table II: LAO vs. STO BE
shift). If band bending is present, such a small difference is
unable to be the sole cause of the build-up of the 2DEG.

\begin{table*}
\centering \caption{
Ti$^{3+}$/Ti$^{4+}$ peak area ratio, energy shift $\Delta$(BE)
between the Ti$^{4+}$-Ti$^{3+}$ core level peaks, Ti 2p and Sr 3d
FWHM, Ti 2p peak broadening ($\sigma_{EXTRA}$) with respect to the
STO case, sheet charge density (SCD) evaluated assuming a Ti$^{3+}$
distribution across a 1 u.c. or a 2 u.c. thick layer below the
interface } \label{tab2}

\begin{tabular}[c]{ccccccccc}
\hline
\hline      $ $     & P(O$_{2}$)       & Ti$^{3+}$/Ti$^{4+}$ & $\Delta$(BE)        &Ti2p      &  Sr 3d    & $\sigma_{EXTRA}$ &    SCD      & SCD    \\
       $ $          & mbar             & XPS                 & Ti$^{4+}$-Ti$^{3+}$ &FWHM      &  FWHM      & (Ti 2p) &  1 u.c.     & 2 u.c. \\
       $ $          & x 10$^{-3}$      & ratio               & (eV)                &(eV)      &  (eV)      & (eV)&          &     \\
$ $                 &                  &                     &                     &          &            &  &             &               \\
\hline 5 u.c. cond. & 1.0              & 0.056$\pm0.005$      & 2.03                & 1.47     &     1.24   &0.56&1.50x10$^{14}$&1.68x10$^{14}$ \\
\hline 3 u.c. ins.  & 1.0              & 0.012$\pm0.005$      & 2.10                & 1.43     &     1.18   &0.44&3.70x10$^{13}$&4.00x10$^{13}$ \\
\hline 5 u.c. ins.  & 100              & 0.006$\pm0.005$      & 1.78                & 1.41     &     ---    &0.37&        ---   &     \\
\hline $STO$        & ---              & 0.004$\pm0.005$      & 1.73               & 1.36     &     1.05   & 0.00&       ---   &     \\
\hline \hline\\

\end{tabular}

\end{table*}

On the intensity side, the accuracy of the fitting results can be
checked by evaluating if the STO signal attenuation obtained from
the fitting is consistent with the thickness of the LAO overlayer.
Defining the integrated intensity of the XPS signal referred to STO
as $I_{STO}$, then the STO signal attenuation of the 5 u.c.
(thickness = d$_5$) sample with respect to the 3 u.c. (d$_3$) sample
can be written as

\begin{equation}
\frac{I_{STO 5 u.c.}}{I_{STO 3 u.c.}}=
\frac{\int_{d_5}^{+\infty}\Phi_{LAO_{d5}-STO}(E_k,\alpha,z)dz}{\int_{d_3}^{+\infty}\Phi_{LAO_{d3}-STO}(E_k,\alpha,z)dz}
\end{equation}

while this ratio is experimentally determined to be about 0.59 for
the two sample grown at low P(O$_2$). Solving the equation to
extract the thickness difference between the two layers, a value of
$d_5-d_3=9$ \r{A} is obtained, in good agreement with the expected
5-3=2 u.c. thickness (7.6 - 7.8 \r{A}). This makes the signal
attenuation evaluated from the VB and shallow core levels
consistent, providing also a positive feedback on the results about
the negligible LAO-STO valence band offset among the samples. It is
also worth noting that the measured peak S$_{III}$, ascribed to La
5p$_{3/2}$, in the 5 u.c. conducting sample (Figure 2-d) is lower
than the calculated peak. This is consistent with the fact the this
sample has been evaluated as La poor from the shallow core level
analysis (Fig.1). Finally, we observe that the use of the inelastic
mean free path alone rather than the DDF, would yield a
$d_5-d_3=13.3$ \r{A} value, clearly in contrast with the expected
thickness. This further shows the inadequacy of the IMFP concept in
the estimation of attenuation lengths.

\subsection{Binding Energy shifts and band offsets}

Possible band offsets can also be measured by considering the BE
difference between a core line from a LAO element and a core line of
an STO element. Following Ref. \cite{LAOSTO_chambers}, we have
evaluated the following differences (Eq. 3 to 8):

\begin{align}
\Delta \text{E}_{V}^{La4d-Sr3d} =\ &(\text{E}_{La4d}-\text{E}_{V})_{LAO} - (\text{E}_{Sr3d}-\text{E}_{V})_{STO} \nonumber \\
&+ (\text{E}_{Sr3d}-\text{E}_{La4d})_{HJ} \\
& \nonumber \\
 \Delta \text{E}_{V}^{Al2s-Sr3d} =\
&(\text{E}_{Al2s}-\text{E}_{V})_{LAO}
- (\text{E}_{Sr3d}-\text{E}_{V})_{STO} \nonumber \\
&+ (\text{E}_{Sr3d}-\text{E}_{Al2s})_{HJ} \\
& \nonumber \\
 \Delta \text{E}_{V}^{La4d-Ti3p} =\
&(\text{E}_{La4d}-\text{E}_{V})_{LAO} -
(\text{E}_{Ti3p}-\text{E}_{V})_{STO} \nonumber \\
&+ (\text{E}_{Ti3p}-\text{E}_{La4d})_{HJ} \\
& \nonumber \\
\Delta\text{E}_{V}^{Al2s-Ti3p} =\
&(\text{E}_{Al2s}-\text{E}_{V})_{LAO} -
(\text{E}_{Ti3p}-\text{E}_{V})_{STO} \nonumber \\
&+ (\text{E}_{Ti3p}-\text{E}_{Al2s})_{HJ}
\end{align}

\begin{align}
\Delta \text{E}_{V}^{Al2p-Ti3p} =\
&(\text{E}_{Al2p}-\text{E}_{V})_{LAO} -
(\text{E}_{Ti3p}-\text{E}_{V})_{STO} \nonumber \\
&+ (\text{E}_{Ti3p}-\text{E}_{Al2p})_{HJ}\\
& \nonumber \\
\Delta \text{E}_{V}^{Al2p-Sr3d} =\
&(\text{E}_{Al2p}-\text{E}_{V})_{LAO} -
(\text{E}_{Sr3d}-\text{E}_{V})_{STO} \nonumber \\
&+ (\text{E}_{Sr3d}-\text{E}_{Al2p})_{HJ}
\end{align}

that can be identified by referring to the diagram of Fig.3, where
the energy levels involved in the calculations for both the
reference LAO and STO single crystals and the LAO-STO
heterostructure are shown. The results are reported in Table II. As
can be observed, when averaged over the six different combinations,
all differences are within 0.15 eV (Table II: ave. core), in
agreement with the results obtained from the VB data analysis (0.21
eV; Table II: LAO vs. STO BE shift).

Several studies on the band lineup have been published so far, but
it is rather difficult to find a rationale among them. A first group
of studies is focussed on core level differences between p-type and
n-type samples. Takizawa \emph{et al.} \cite{fuji_2011} have
focussed on the band lineup of p-type and n-type interfaces, both
with 1,3,4,5, and 6 LAO u.c. overlayers. They show that core levels
belonging to the same layer of the HS (i.e. La and Al for LAO, and
Ti and Sr for STO) do not show appreciable shifts (less than 100
meV). In turn, an energy shift is observed for the Al2p-Sr3d BE
difference. This shift increases with the number of LAO u.c. for
both p and n-type samples. The interfaces have been grown by PLD in
a 1x10$^{-5}$ torr oxygen partial pressure. Segal \emph{et al.}
\cite{segall} present a similar study, where p-type and n-type
interfaces with 2 up to 9 LAO u.c. have been investigated and the
Sr3d-La4d BE difference has been tracked. Here, this energy
difference is found to change with the number of layers but while it
increases for p-type samples, it decreases for n-type samples. The
samples have been grown by MBE in 3x10$^{-7}$ torr oxygen partial
pressure. Yoshimatsu \emph{et al.} \cite{yoshi08} investigate the Ti
2p BE shift for both p-type and -type samples (0 to 6 LAO u.c.,
LPLD, $1x10^{-5}$ torr). In this case, the Ti 2p BE for the p-type
samples is found to be constant, while for the n-type samples the BE
decreases with LAO thickness.

Unlike the papers so far mentioned, Chambers \emph{et al.}
\cite{LAOSTO_chambers} did not consider the p-type and n-type set of
samples, but focussed on two 4 u.c. n-type samples grown at rather
similar O$_{2}$ pressures (1x10$^{-5}$ Torr and 8x10$^{-6}$ Torr). A
set of BE differences have been evaluated, namely Sr 3d - La 4d, Sr
3d- Al 2p, Ti 2p - La 4d, and Ti 2p - Al 2p. Furthermore these
differences were referenced to the VBO, a procedure neglected in the
previous studies. The present study follows this pathway, but
important differences emerge with respect to Chamber's results. It
is rather interesting to note that in the samples considered in
Ref.\cite{LAOSTO_chambers} the Ti$^{3+}$ contribution is not
detected on the low-BE of the Ti 2p$_{3/2}$ XPS peak. In the samples
presented here we detect Ti$^{3+}$ states in Ti 2p core lines, which
find a counterpart in the 3d$^{1}$ electron emission in the valence
band region already probed by resonant photoemission (RESPES) in a
previous investigation \cite{APL_LAOSTO}. Furthermore, while
Chambers detects similar BE energy differences in the same sample,
irrespectively of the couple of atoms selected, we detect much
different changes within the same sample. When Ti is involved as one
of the core levels in Eq. 5-7, larger shift are usually detected
with respect to those resulting from Eq.3, 4, and 8. This seems to
indicate that in our sample it is possible to observe element
specific shifts, while in the samples examined by Chambers the shift
are rather uniformly distributed. This is tentatively ascribed to
the fact that localized Ti$^{3+}$ states at the interface may have
much larger effects on the local electronic properties than the
widespread charge distribution invoked by Chambers to justify the
lack of Ti$^{3+}$ low-BE feature in Ti 2p XPS spectra.

\begin{table*}
\centering \caption{Calculated BE shift between core level peaks,
calculated according to Eqs. 3 to 8, LAO vs STO VB spectra energy
shift obtained by the fitting of the LAO-STO VB spectra. All
energies in eV. } \label{tab222}

\begin{tabular}[c]{ccccccccc}
\hline
\hline      $ $       & La4d$_{5/2}$-Sr3d    & Al2s-Sr3d & La4d$_{5/2}$-Ti3p & Al2s-Ti3p & Al2p-Ti3p  & Al2p-Sr3d   &ave.               & LAO vs STO  \\
       $ $       &              &           &           &           &            &                             &core               &  VB shift   \\
\hline 5 u.c. cond.        & -0.46        & -0.54     & -0.89     & -0.97    &   -0.94    &   -0.52     & -0.72 $\pm$ 0.24     & -0.04 \\
\hline 3 u.c. ins.         & -0.56        & -0.61     & -0.74     & -0.79    &   -0.75     &  -0.57     & -0.67 $\pm$ 0.10     & -0.10 \\
\hline 5 u.c. ins.      & -0.34        & -0.27     & -1.01     & -0.94    &   -0.77     &  -0.10     & -0.57 $\pm$ 0.38     & +0.17 \\
\hline
\end{tabular}

\end{table*}

\subsection{Ti 2p core levels}

In Fig.\ref{fig_XPS_Ti2p} the Ti 2p XPS core lines of the 3 u.c.,
the two 5 u.c. heterointerfaces and the pure STO are shown. The Ti
2p spectrum is almost identical to that expected for a Ti$^{4+}$
ion. However, a small bump is detectable on the low BE side of the 5
u.c. spectrum, in the position usually associated to Ti$^{3+}$
electronic states. These states can be detected only through a
comparison among different samples and can be easily confused as an
additional experimental broadening of Ti 2p$_{3/2}$ peak. For this
reason we also show the Ti 2p core line of SrTiO$_{3}$, measured
with the same energy resolution. The Ti 2p$_{3/2}$ peak FWHM of the
5 u.c. conducting compound (1.47 eV) is larger than those of the 5
u.c. (1.41 eV) and the 3 u.c. (1.43 eV) insulating compounds. It is
important to note that the narrower Ti 2p peak is that of STO (1.36
eV). Therefore, we see an overall decrease of the Ti 2p FWHM from
the conducting LAO-STO to the insulating STO single crystal. We
exclude broadening effects due to charging, which should be opposite
to those observed, and therefore we regard the broadening as due to
an intrinsic effect. Also the peak area of Ti$^{3+}$ states of the
conducting compound is larger than that of the two insulating
compounds. The Ti$^{3+}$/Ti$^{4+}$ ratio sharply decreases from
0.056 (5 u.c.) to 0.012 (3 u.c.) and is nearly negligible for the 5
u.c. insulating sample (0.006) and the STO single crystal (0.004),
assuming an uncertainty of $\pm$0.005 on the peak ratio values.
Furthermore, also the Ti$^{3+}$ BE of the samples grown in low
P(O$_{2}$) is different from that measured for the other samples,
the insulating 5 u.c. and the reference STO single crystal. This
peak is found about 2 eV below the main line, but this difference is
reduced when the  5 u.c. insulating (1.78 eV) and the STO (1.73 eV)
samples are considered, indicating a different origin for these
peaks. Indeed, an insulating sample has been reported\cite{KOI_2011}
to show a Ti$^{3+}$ peak closer to the main line with respect to a
conducting LAO-STO interface.

\begin{figure}
\begin{center}
\includegraphics[ width=0.43\textwidth]{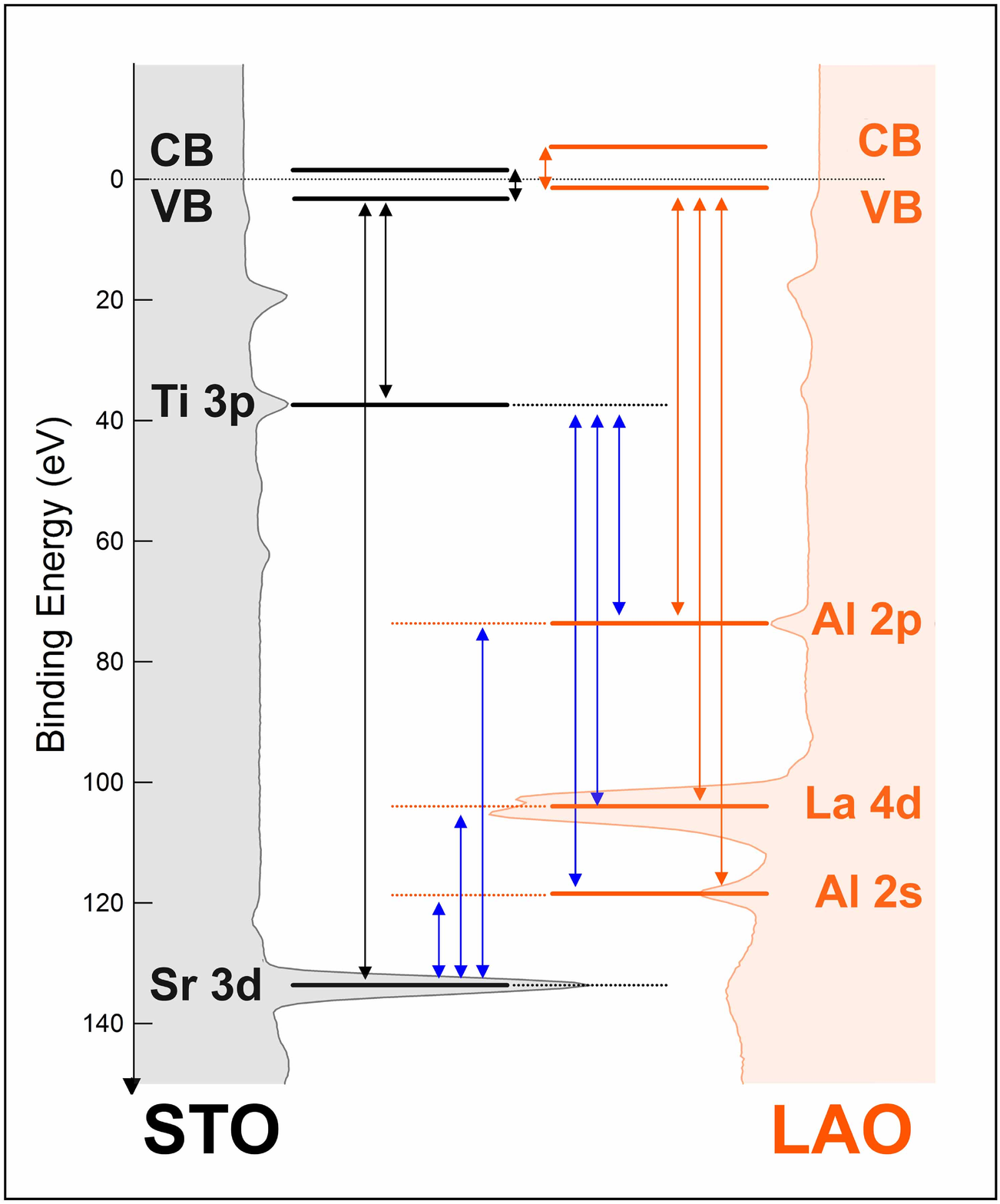}
\caption{(Color online) Schematic of energy levels involved in the
band offset calculations. The STO core levels are reported on the
left side, while the LAO core levels can be found on the right side.
The shaded areas show the XPS data collected from the STO and LAO
bulk crystals.}
\end{center}
\end{figure}

The presence of Ti$^{3+}$ states was controversial in early studies
on LAO-STO, as not all authors observed these features in spite of
the conducting nature of their samples. As already mentioned, a
signature of Ti$^{3+}$ states is the feature appearing on the low BE
side of the Ti 2p$_{3/2}$ core line. In addition, this feature
should have a counterpart in the valence band region, detectable
either through RESPES at the Ti L-edge, or through UPS photoemission
\cite{siemo07}. While Takizawa \cite{fuji_2011}and Yoshimatsu
\cite{yoshi08}discuss the properties of Ti 2p core lines, in Segal's
study \cite{segall} no mention is done on Ti 2p. Yoshimatsu \emph{et
al.} do not report on the Ti$^{3+}$ contribution, but they focus on
the BE shift of the Ti 2p$_{3/2}$ core line, while Takizawa \emph{et
al.} discuss the Ti$^{4+}$/Ti$^{3+}$ ratio, but do not report on Ti
2p$_{3/2}$ BE shift. Furthermore, unlike more recent
studies\cite{KOI_2011, APL_LAOSTO}, Yoshimatsu \emph{et al.} do not
find any evidence of 3d$^{1}$ states in RESPES experiments at the Ti
L-edge resonance. For the present samples, evidence of Ti 3d states
in the gap are provided in Ref. \cite{drera2012}, Par.5.4.2.

In addition to Ti 2p, also the Sr 3d core levels of the
heterostructures appear to be different from those of STO (Figure
\ref{fig_XPS_SrAlLa}-c). As shown in Table II, in spite of the high
quality of the STO side of the heterojunction, the Sr peaks are not
as sharp as pristine STO. The Sr 3d peak of the conducting 5 u.c.
sample 1.24 eV) is larger than that of the corresponding 3 u.c.
layer 1.18 eV), both being larger than the peak measured for STO
1.05 eV). This trend is consistent with that measured for the Ti 2p
core lines, suggesting the presence of disorder (cationic exchange
or oxygen vacancies) around strontium atoms at the interface. This
structural disorder is supposed to alter the ideal structural
environment around the Ti and Sr cations, yielding potential
fluctuations that ultimately may results in a Ti 2p and Sr 3d peak
width broadening. Similar effects have been observed in, e.g., the
Ti 2p core lines of Fe-doped rutile single crystals
\cite{ironrutile}.

An alternative picture to be considered, that is however supported
by a qualitative analysis, is based on possible band bending effects
on the peak width. Segal \emph{et al} have considered this
hypothesis and they have been able to estimate a band bending
smaller than that expected from theoretical predictions. Indeed,
Segal \emph{et al} have investigated these effects on the La
4d$_{3/2}$ peak width, obtaining quite smaller broadening with
respect to those expected from the polar catastrophe theory.
Furthermore they do not find a specific trend in the FWHM, as the
data appear to be scattered.

In the present study we choose to address this question on the
elements (Sr and Ti) of the buried interface. Assuming that the
increase of peak width is due to band bending, it is interesting to
relate the observed FWHM with the VBO by comparing the results of
Table I and Table II. Here, the Ti 2p peak width in STO (1.36 eV) is
regarded as the width of the "ideal" Ti-terminated STO without the
LAO capping layer. Based on this, we can extract the extra width (
$\sigma_{EXTRA}$) due to the interface effects on the basis of the
following Equation: $\sigma_{TOTAL}=(\sigma_{STO}^2+
\sigma_{EXTRA}^2)^{1/2}$. The results are reported in Table I. As
can be observed, $\sigma_{EXTRA}$  shows a decrease similar to that
found for the average core level shift that appears in Table II.
This indicates that larger average core level shifts yield larger Ti
2p peak widths.

\begin{figure}
\begin{center}
\includegraphics[width=0.45\textwidth]{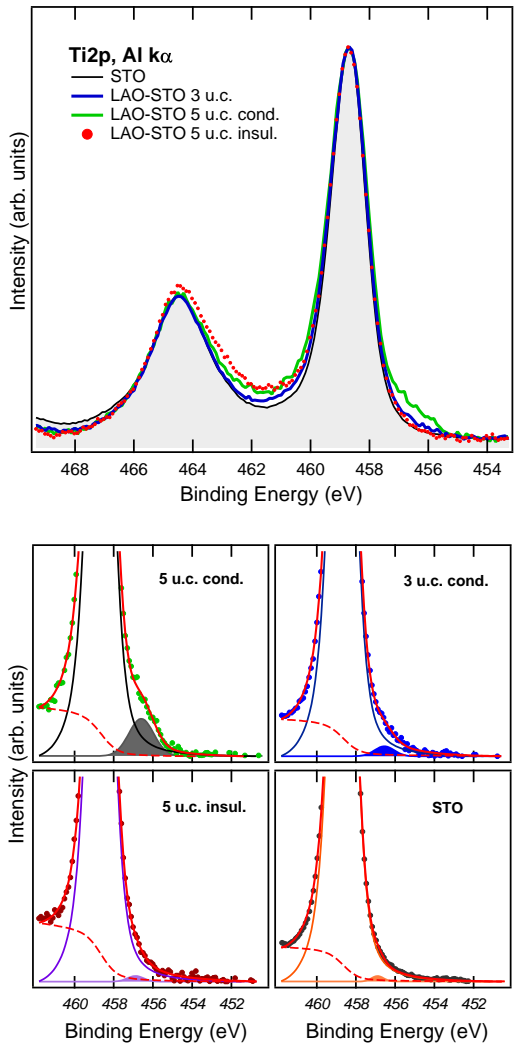}
\caption{(Color online) Top panel: Ti 2p core level XPS spectra
collected from the 3 u.c insulating HS (thick line), the 5 u.c.
conducting HS (thin line), the 5 u.c. insulating HS (dots), and the
STO single crystal (shaded area). Bottom panels: Fitting (thin
lines) of the Ti 2p$_{3/2}$ core levels (dots) with two peaks,
accounting for the Ti$^{4+}$ and the Ti$^{3+}$ (shaded area)
contributions, and an integral (Shirley type, dashed line)
background. \label{fig_XPS_Ti2p}}
\end{center}
\end{figure}

Finally, from the Ti$^{3+}$/Ti$^{4+}$ intensity ratio, by properly
considering the attenuation of the Ti signal due to the LAO
overlayer, an estimation of the sheet charge density (SCD) can also
be provided.

\begin{figure*}
\begin{center}
\includegraphics[width=0.40\textwidth]{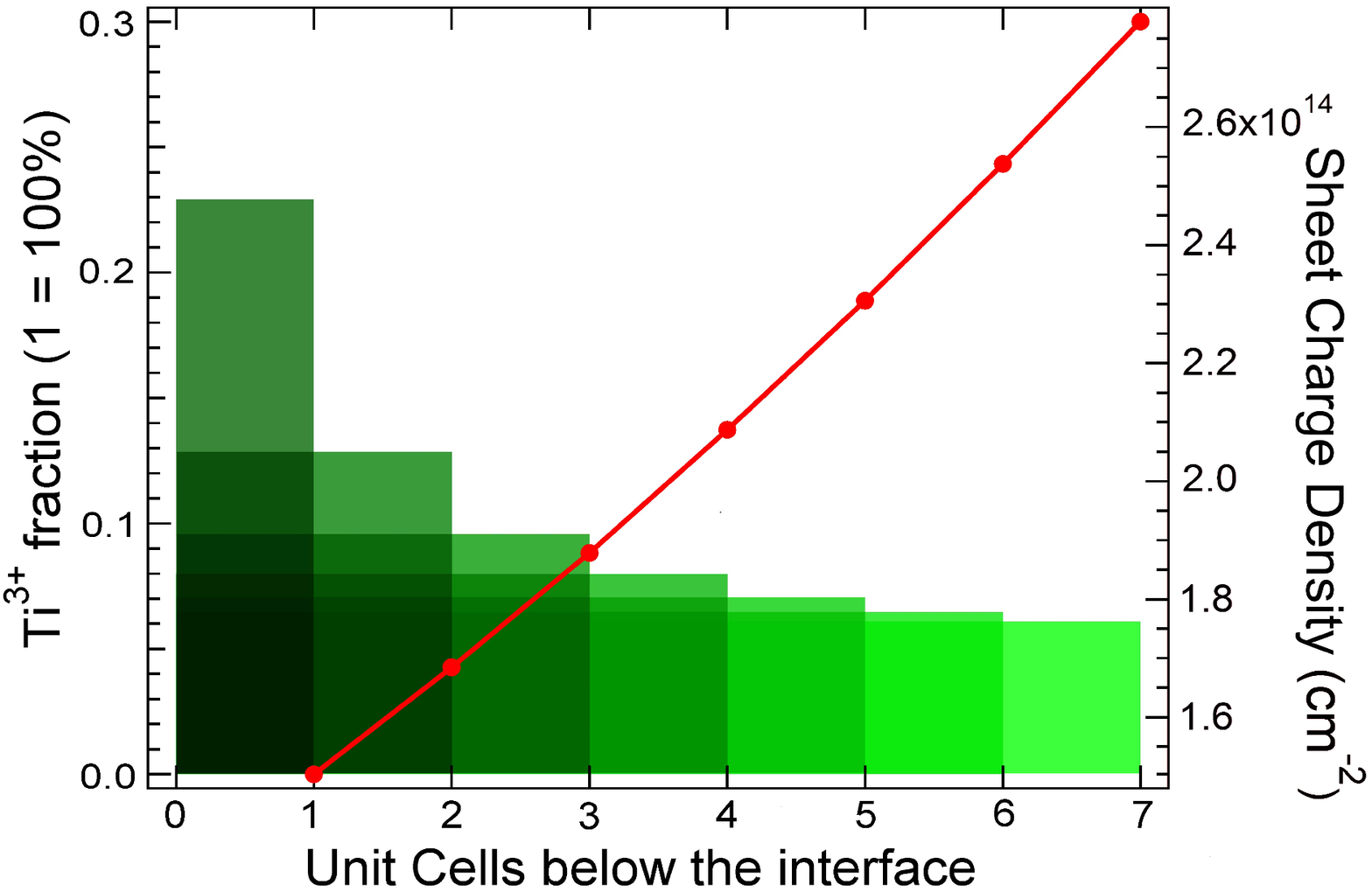}
\includegraphics[width=0.40\textwidth]{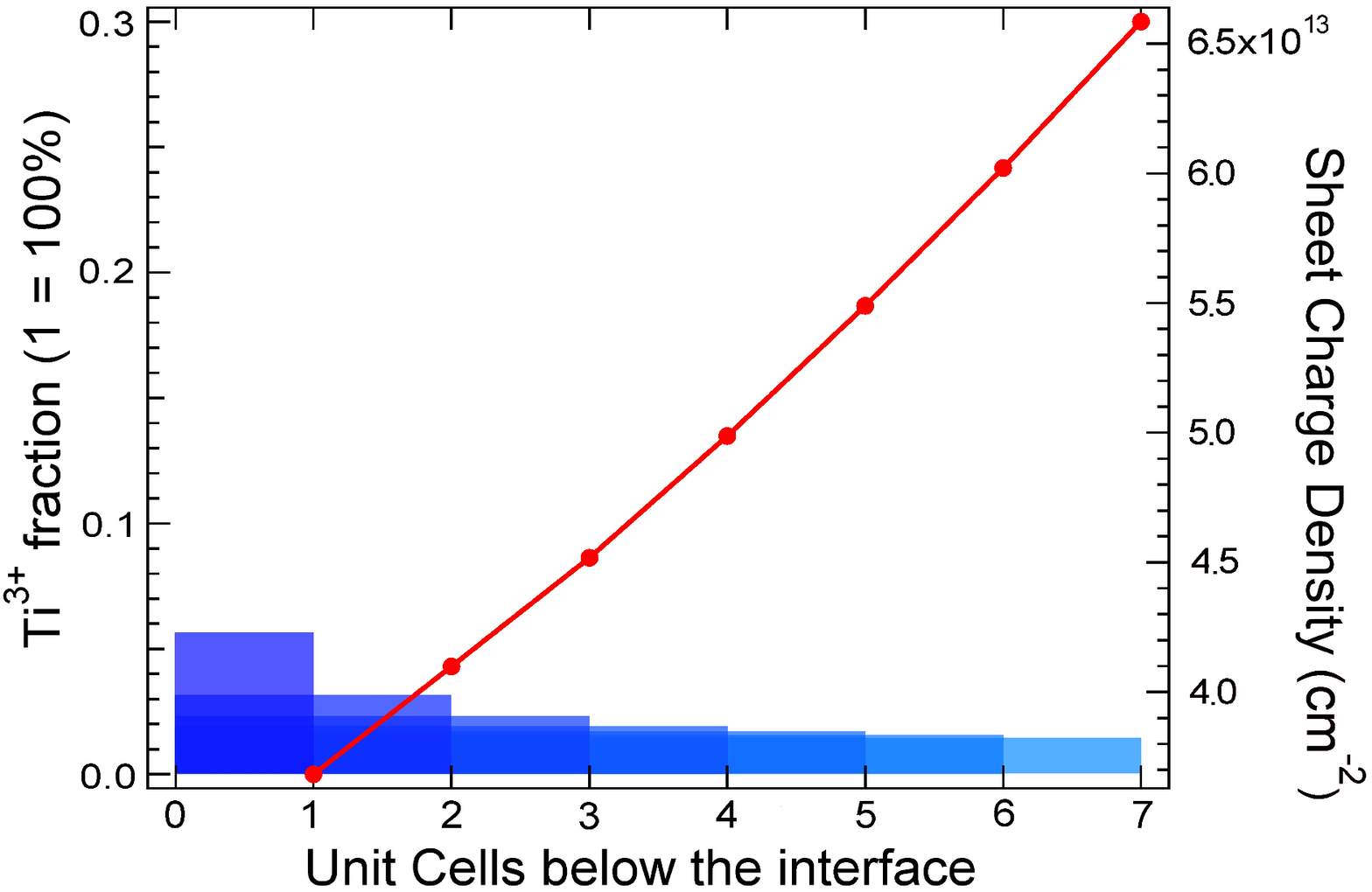}
\caption{(Color online) Charge profile and corresponding sheet
charge density (SCD) for the 5 u.c. sample (left panel) and the 3
u.c. sample (right panel). \label{charge}}
\end{center}
\end{figure*}

The evaluation of the Ti$^{3+}$/Ti$^{4+}$ ratio has been drawn on
the basis of the DDF concept, using the XPS peak area values
reported in Table I. The results are shown in Fig.\ref{charge}. Our
calculations are based on the fact that the Ti$^{3+}$/Ti$^{4+}$
ratio reported in Table I can be generated by different
distributions of the Ti$^{3+}$ ions below the interface, e.g. the
same ratio can be produced by a high density of ions close to the
interface or by a low density of ions distributed across a larger
layer below the interface. We start our analysis by assuming that,
once the width of the charge profile is established, the Ti$^{3+}$
ion distribution is constant below the interface. A different
profile, e.g. a charge profile decaying with depth below the
interface, can be ultimately described by a linear combination of
constant charge profiles, as our calculation is carried out on the
quite fine discrete steps (i.e. the thickness of a unit cell).

In Fig.\ref{charge} the height of the histogram bars represent the
Ti$^{3+}$ fraction predicted for a uniform distribution of Ti$^{3+}$
atoms in a layer below the interface as thick as the width of the
histogram bar. The thickness of this layer is assumed to be a
multiple of the STO unit cell parameter. The corresponding sheet
charge density (SCD) is evaluated for each histogram bar and is
shown as a thick line (right axis). Assuming that al the Ti$^{3+}$
ions are located in the first unit cell below the interface, the
resulting SCD is 1.5x10$^{14}$ cm$^{-2}$. This value represents the
lowest limit for the SCD calculated on the basis of the experimental
Ti$^{4+}$/Ti$^{3+}$ XPS peak area ratio.

A much lower estimation of the sheet charge density is drawn for the
insulating 3 u.c. sample. In this case, the lowest limit for the SCD
is 3.7x10$^{13}$ cm$^{-2}$. This value can be regarded as the
intrinsic density of carries at the interface for the two samples
grown al low P(O$_2$), which is found independently on the physical
mechanisms at the basis of the 2DEG build-up. If we consider a
constant charge density spread  about 2 u.c. below the interface,
the present results are in good agreement with those reported by
Sing \emph{et al.} on the basis of HAXPES experiments
\cite{LAOSTO_haxpes}. In both cases, the estimated SCD of the
conducting interface (2x10$^{14}$ ) is higher than that typically
obtained from transport measurements on conducting LAO-STO
interfaces (about 2-6 x10$^{13}$) \cite{LAOSTO_science, dubroka}.
The SCD resulted to be high also for the 3 u.c. sample (about
4x10$^{13}$). This discrepancy seems to point out that only a
fraction of the Ti$^{3+}$ states detected by XPS contribute to the
2DEG. One explanation can be found in a Ref.\cite{seo}, where a
distinction between two kind of charge carriers is provided: low
density high mobility carriers for the transport measurements and
high density low mobility carriers from optical measurements.
Alternatively, the formation of photoinduced charge carriers either
by X-ray or ambient light irradiation has to be considered. However,
under this assumption, it is not straightforward to explain the lack
of such large photoinduced effects in the insulating 3 u.c. sample
and also in STO. In particular, as these effects should occur on the
STO side of the junction to yield Ti$^{3+}$ electronic states, it is
difficult to justify the lower density for the 3 u.c. sample, where
irradiation effects are supposed to be larger than in the 5 u.c.
sample due to the thinner LAO overlayer on top of the STO substrate.
Finally we observe that an estimate of Ti$^{3+}$ fraction carried
out with the IMFP attenuation length alone, without considering the
full DDF approach, would overestimate the Ti$^{3+}$ fraction, as
neglecting, e.g., the elastic scattering events, the resulting
attenuation length is higher. Namely, by using the IMFP for STO
($\lambda$=21.67 ${\AA}$ at KE= 1000 eV) we would obtain a Ti$^{3+}$
fraction of 0.34 rather than 0.23 for a distribution depth of 1
u.c., and 0.19 rather than 0.12 for a distribution depth of 2 u.c.,
i.e. an overall overestimation of about 50 percent. This would yield
an analogous overestimation of the SDC.

\section{Conclusions}

Far from being thoroughly assessed in the literature, the
spectroscopic signature of 2DEG in LAO-STO interfaces is discussed
in the present study, where we add novel data to the set of
experiments so far reported. Indeed, rather than focussing on the
difference between p-type and n-type samples, we choose to focus on
oxygen stoichiometry effects and on the comparison with parent LAO
and STO compounds. Unlike previous studies, we show that in LAO-STO
heterostructures (i) the Ti$^{3+}$ contribution to Ti 2p core
levels, (ii) the Ti 3d$^{1}$ in-gap states \cite{drera2012}, and
(iii) larger Ti 2p width with respect to STO, can be simultaneously
present, though at different extent, suggesting that all these
signatures of the 2DEG are accessible via photoemission on each
sample.

Three LAO-STO interfaces have been analyzed by X-ray photoemission
spectroscopies, plus a LAO and STO reference single crystals. The
energy, width and intensity of core level peaks, and the valence
band spectra have been carefully considered in order to look for
band-bending effects at the heterointerface and probe the dependence
of the Ti$^{3+}$ charge density on the growth conditions.

In the analysis of VBO, we consider the approach suggested by
Chambers \emph{et al.} \cite{LAOSTO_chambers} by referencing the BE
also to separated LAO and STO valence band maxima. Unlike the
reported findings, we find BE differences sensitive to the choice of
elements, which suggests that different atoms in the interface may
undergo different energy shifts. In particular, when Ti is involved
the major differences are estimated.

From the analysis of core level energies, we exclude the presence of
band bending effects larger than about 0.7 eV, ruling out tunneling
from the LAO offset valence band to STO empty 3d levels as the sole
mechanism for the build-up of the 2DEG. Differences in band offset
among the three samples are quite limited, spanning a range of about
0.2 eV, in spite of the remarkably different growth conditions and
electrical properties. Likewise, differences among relative BE shift
required to best-fit the LAO-STO valence band are small. However, a
correlation among these differences, the peak width of Ti and Sr at
the interface, and band bending is found, showing that band offset
is at work to shape the energy landscape and that it is possible to
consistently single out these effects by a careful analysis of
spectroscopic data. Indeed,the FWHM of the Ti 2p and Sr 3d core
lines is larger for the conducting sample with respect to any of the
insulating interfaces and the STO single crystal. The intrinsic
origin of this width is ascribed to band bending effects, though
disorder effects (electronic or structural) around the photoemitting
atom in the conducting sample may by at work at the same time.

Finally, we have shown that the density of Ti$^{3+}$ levels strongly
depends both on the LAO overlayer thickness and oxygen partial
pressure during the growth. Two heterostructures grown at the same
P(O$_{2}$) present a well detectable Ti$^{3+}$ peak about 2 eV below
the Ti$^{4+}$ main line. The STO and the 5 u.c. insulating samples
show a much weaker contribution of Ti$^{3+}$ states, with also a
different binding energy. On the basis of the Ti$^{3+}$/Ti$^{4+}$
XPS peak area ratio, a SCD larger than that expected from transport
measurements has been evaluated for the two samples grown at low
P(O$_{2}$), the SCD of the 5 u.c. layer being about an order of
magnitude larger than in the 3 u.c. sample.

\section{Acknowledgments} Support from the Dutch FOM and NWO
foundations is acknowledged.


\begin {thebibliography} {apssamp}


\bibitem{LAOSTO_ohtomo}
H. Y. Hwang and A. Ohtomo,  Nature \textbf{427}, 423 (2004).

\bibitem{LAOSTO_science}
S. Thiel, G. Hammer, A. Schmehl, C. W. Schneider, J. Mannhart,
Science \textbf{313} (2006), \textbf{1942}

\bibitem{LAOSTO_rev1}
R. Pentcheva and W. E. Pickett, J. Phys.: Cond. Mat. \textbf{22},
043001 (2010).

\bibitem{polarcatastrophe1}
N. Nakagawa, H. Y. Hwang, and D. A. Muller, Nature Mater.
\textbf{5}, 204 (2006).

\bibitem{polarcatastrophe2}
J. Gonjakowski, F. Finocchi, and C. Noguera, Rep. Prog. Phys.
\textbf{71}, 016501 (2008).

\bibitem{LAOSTO_rev2}
N. Reyren, S. Thiel, A. D. Caviglia, L. Fitting Kourkoutis, G.
Hammerl, C. Richter, C. W. Schneider, T. Kopp, A. S. Rüetschi, D.
Jaccard, M. Gabay, D. A. Muller, J. M. Triscone, and J. Mannhart,
Science \textbf{317}, 1196 (2007).

\bibitem{LAOSTO_rev3}
A. Brinkman, M. Huijben, M. van Zalk, J. Huijben, U. Zeitler, J.K.
Maan, W.G. van der Wiel, G. Rijnders, D.H.A. Blank and H.
Hilgenkamp, Nature Materials \textbf{6}, 493 (2007).


\bibitem{LAOSTO_chambers}
S. A. Chambers, M. H. Englehard, V. Shutthanandan, Z. Zhu, T. C.
Droubay, T. Feng, H. D. Lee, T. Gustafsson, E. Garfunkel, A. Shah,
J. M. Zuo, and Q. M. Ramasse, Surf. Sci. Rep. \textbf{65}, 317
(2010).

\bibitem{VONK}
V. Vonk, J. Huijben, D. Kukuruznyak, A. Stierle, H. Hilgenkamp, A.
Brinkman, and S. Harkema, Phys. Rev. B \textbf{85}, 045401  (2012)

\bibitem{BandBending}
R. Pentcheva and W. E. Pickett, Phys. Rev. Lett. \textbf{102},
107602 (2009).

\bibitem{OxyVac1}
A. S. Kalabukhov, \emph{et al.}, Phys. Rev. Lett. \textbf{103},
146101 (2009)

\bibitem{OxyVac2}
G. Herranz, \emph{et al.}, Phys. Rev. Lett. \textbf{98}, 216803
(2007).

\bibitem{OxyVac3}
W. Siemons, \emph{et al.}, Phys. Rev. Lett. \textbf{98}, 196802
(2007).

\bibitem{OxyVac4}
M. Basletic, \emph{et al.}, Nat. Matter. \textbf{7}, 621-625 (2008).

\bibitem{LAOSTO_natmat}
G. Rijnders and D. H. A. Blank, Nature Materials \textbf{7}, 270
(2008).

\bibitem{APL_LAOSTO}
G. Drera, F. Banfi, F. Federici Canova, P. Borghetti, L. Sangaletti,
F. Bondino, E. Magnano, J. Huijben, M. Huijben, G. Rijnders, D. H.
A. Blank, H. Hilgenkamp, and A. Brinkman, Appl. Phys. Lett.
\textbf{98}, 1 (2011).

\bibitem{KOI_2011}
A. Koitzsch, J. Ocker, M. Knupfer, M. C. Dekker, K. D\"{o}rr, B.
Buchner, P. Hoffmann, Phys. Rev. B \textbf{84}, 245121 (2011)

\bibitem{fuji_2011}
M. Takizawa, S. Tsuda, T. Susaki, H. Y. Hwang, and A. Fujimori,
Phys. Rev. B, \textbf{84}, 245124 (2011)

\bibitem{LAOSTO_haxpes}
M. Sing, G. Berner, K. Goss, A. Muller, A. Ruff, A. Wetscherek, S.
Thiel, J. Mannhart, S. A. Pauli, C. W. Schneider, P. R. Willmott, M.
Gorgoi, F. Schafers, and R. Claessen, Phys. Rev. Lett. 102, 176805
(2009).

\bibitem{segall}
Y. Segal, J.H. Ngai, J. W. Reiner, F. J. Walker, C. H. Ahn, Phys.
Rev. B \textbf{80}, 241107 (2009)

\bibitem{WernerDDF}
W. S. M. Werner, Surf. Int. Anal. \textbf{31}, 141 (2001).

\bibitem{JablonskiTA}
A. Jablonski, Phys. Rev. B \textbf{58}, 16470 (1998).

\bibitem{TilinDDF} I.S. Tilinin, A. Jablonski, J. Zemek ,S. Hucek
J. Electr. Spec. Rel. Phen. \textbf{97}, 127-140 (1997)

\bibitem{x_section}
J. J. Yeh and I. Lindau, Atomic Data and Nuclear Data Tables
\textbf{32}, 1-155 (1985).

\bibitem{QIAO2011}
L. Qiao, T.C.Droubay, T. Varga, M.E. Bowden, V. Shutthanandan, Z.
Zhu, T.C. Kaspar, S.A. Chambers, Phys. Rev. B \textbf{83}, 085408
(2011)

\bibitem{La2O3}
M.F. Sunding, K. Hadidi, S. Diplas, O.M. L${\O}$vvik, T.E. Norby,
A.E. Gunn${\ae}$s, Journ. of Electr. Spectr. and Rel. Phen.
\textbf{184}, 399 (2011).

\bibitem{yoshi08}
K.Yoshimatsu, R. Yasuhara, H. Kumigshira, and M. Oshima, Phys. Rev.
Lett. \textbf{101 }, 026802 (2008)

\bibitem{siemo07}
W. Siemons, G. Koster, H. Yamamoto, T. H. Geballe, D. H. A. Blank,
M. Beasley, Phys. Rev. B \textbf{76}, 155111 (2007)

\bibitem{drera2012}
G. Drera, arXiv:1210.8000 [cond-mat.str-el]

\bibitem{ironrutile}
L. Sangaletti, M.C. Mozzati, G. Drera, P. Galinetto, C.B. Azzoni, A.
Speghini, M. Bettinelli, Phys. Rev. B \textbf{78}, 075210 (2008)

\bibitem{dubroka}
A. Dubroka \emph{et al.}, Phys. Rev. Lett. \textbf{104}, 156807
(2010)

\bibitem{seo}
S. S. A. Seo, Z. Marton, W. S. Choi, G. W. J. Hassink, D. H. A.
Blank, H. Y. Hwang, T. W. Noh, T. Egami, and H. N. Lee, Appl. Phys.
Lett. \textbf{95}, 082107 (2009);

\end{thebibliography}
\end{document}